\documentclass [11pt, prd,onecolumn,nofootinbib,notitlepage]{revtex4-1}
\usepackage{amsfonts,amsmath,amscd, amssymb}
\usepackage{graphicx}% Include figure files
\usepackage{caption}
\usepackage{subcaption}
\usepackage{bbm}

\def\A{\mathsf{A}}
\def\AN{\mathsf{AN}}
\def\an{\mathsf{an}}

\def\F{\mathsf{F}}

\def\SO{\mathsf{SO}}
\def\Ad{\mbox{Ad}}

\def\so{\mathsf{so}}

\begin{document}

\title{Deformed Carroll particle from 2+1 gravity}

\author{Jerzy Kowalski-Glikman}
\email{jerzy.kowalski-glikman@ift.uni.wroc.pl}\affiliation{Institute
for Theoretical Physics, University of Wroc\l{}aw, Pl.\ Maksa Borna
9, Pl--50-204 Wroc\l{}aw, Poland}
\author{Tomasz Trze\'{s}niewski}
\email{tomasz.trzesniewski@ift.uni.wroc.pl}\affiliation{Institute
for Theoretical Physics, University of Wroc\l{}aw, Pl.\ Maksa Borna
9, Pl--50-204 Wroc\l{}aw, Poland}

\date{\today}
\small
\begin{abstract}\noindent
We consider a point particle coupled to 2+1 gravity, with de Sitter
gauge group $\SO(3,1)$. We observe that there are two contraction
limits of the gauge group: one resulting in the Poincar\'{e} group,
and the second with the gauge group having the form $\AN(2) \ltimes
\an(2)^*$. The former case was thoroughly discussed in the
literature, while the latter leads to the deformed particle action
with de Sitter momentum space, like in the case of
$\kappa$-Poincar\'{e} particle. However, the construction forces the
mass shell constraint to have the form $p_0^2 = m^2$, so that the
effective particle action describes the deformed Carroll particle.
\end{abstract}
\maketitle

Gravity in 2+1 dimensions seems to be, at first sight, an incredibly
dull theory. It does not possess any local degrees of freedom and
local Newtonian interactions between masses as well as gravitational
waves are absent \cite{Staruszkiewicz:1963zza}, \cite{Deser:1983tn},
\cite{Deser:1983nh}. Therefore it can be described by a topological
field theory as firmly established in the seminal paper of Witten
\cite{Witten:1988hc} (and slightly earlier in
\cite{Achucarro:1987vz}.) Remarkably, the picture changes
dramatically when one adds point particles to pure gravity. Then the
gauge degrees of freedom of gravity at the particle's worldlines
become dynamical. By solving for \cite{Matschull:1997du},
\cite{Meusburger:2003ta} (or integrating out in the path integral
formalism \cite{Freidel:2005bb}, \cite{Freidel:2005me}) the
remaining (gauge) degrees of freedom of gravity we are left with a
nontrivial particles dynamics, which includes not only the ``pure''
particles' degrees of freedom, but also the back reaction resulting
from the presence of the gravitational field created by the particles
themselves. Since in 2+1 dimensions gravitational ``action at a
distance'' is absent, the only thing that this back reaction can do
is to {\em deform} the original free particles' actions.

In this letter we show that one of the possible deformed actions,
which can be obtained from 2+1 gravity is an action of the deformed
Carroll particle with $\AN(2)$ momentum space
\cite{KowalskiGlikman:2003we}, \cite{Arzano:2010kz} and
$\kappa$-Minkowski (noncommutative) spacetime
\cite{Lukierski:1993wx}, \cite{Majid:1994cy}. The Carroll particle
\cite{Bergshoeff:2014jla} is a relativistic particle model in the
limit in which the velocity of light becomes zero. Such a particle
cannot move ({\it \ldots\ it takes all the running you can do, to
keep in the same place \ldots}\cite{Carroll}) and its relativistic
symmetry group is a particular contraction of the Poincar\'e group
\cite{Bacry:1968zf}, \cite{Huang:2010if}. The Carroll group has
attracted some attention recently, because it seems to become
potentially relevant in several distinct fields of theoretical
physics, see \cite{Duval:2014uoa} for discussion and references. It
also arises in loop quantum cosmology \cite{Mielczarek:2012tn} in
the context of so-called asymptotic silence, of which the Carrollian
(or ultralocal) limit is a particular realization. More
specifically, the Carrollian limit appears at the transition between
the low-curvature (Lorentzian) regime and the high-curvature regime
in which the metric is Euclidean. It is supposed that at this
transition the symmetry group should change from Poincar\'e to
Carroll and eventually to Euclidean.

In all these cases a simple free dynamical relativistic particle
model possessing Carroll symmetry is of great interest, because it
exhibits physics of the Carrollian world, providing an insight into
it. It is reasonable to expect that in this world some remnants of
particles' interactions should still be present, and that the
situation is similar to that of gravity coupling in 2+1 dimensions.
Indeed, in the limit $c \rightarrow 0$ the local interactions are
frozen, and only the topological sector of gravity can be present.
This is exactly the situation that we encounter in 2+1 dimensions,
where local degrees of freedom of gravity are not possible.
Therefore it could be claimed that the $D + 1$, $D > 2$ analogue of
the 2+1 deformed particle action, derived below, might be the correct
effective description of the Carrollian particle interacting with its
own gravitational field.

Let us start with a short discussion of 2+1 gravity coupled to
particles. According to \cite{Witten:1988hc} 2+1 gravity can be
described by a Chern--Simons theory with an appropriate gauge group;
here we will use the 2+1-dimensional de Sitter group, with three
Lorentz generators $J_a$ and three translational generators $P_a$,
satisfying
\begin{equation}\label{eq:01}
[J_a,J_b] = \epsilon_{ab}^{\ \ c}\, J_c\,, \quad [J_a,P_b] =
\epsilon_{ab}^{\ \ c}\, P_c\,, \quad [P_a,P_b] =
-\epsilon_{ab}^{\ \ c}\, J_c\,,
\end{equation}
where, since the cosmological constant is absorbed into the
translational generators, all generators are dimensionless. It is
convenient to make use of the time plus space decomposition of
spacetime and to accordingly decompose the Chern--Simons connection
one-form into
\begin{equation}\label{eq:02}
\A = A_0 dt + \A_S\,.
\end{equation}
The Lagrangian of the Chern--Simons theory of connection $\A$ coupled to a point particle takes the form
\begin{equation}\label{eq:03}
L = \frac{k}{4\pi} \int \left<\dot\A_S \wedge \A_S\right> -
\left<{\cal C}\, h^{-1} \dot h\right> +
\int \left<A_0, \frac{k}{2\pi} \F_S -
h {\cal C} h^{-1} \delta^2(\vec{x})\, dx^1 \wedge dx^2\right>\,,
\end{equation}
where the spatial curvature $\F_S = d\A_S + [\A_S,\A_S]$. Let us
pause for a  moment to explain the meaning of the terms in this
Lagrangian. The bracket $\left<\ast\right>$ denotes an Ad-invariant
inner product on the Lie algebra of the gauge group, which in our
case is defined to be
\begin{equation}\label{eq:04}
\left<J_a P_b\right> = \eta_{ab}\,, \quad \left<J_a J_b\right> =
\left<P_a P_b\right> = 0\,.
\end{equation}
The coupling constant $k$ can be related to the physical
parameters, the Planck mass $\kappa$ (which in 2+1 dimensions is
purely classical) and the cosmological constant as follows
\begin{equation}\label{eq:04a}
\frac{k}{2\pi} = \frac{\kappa}{\sqrt\Lambda}\,.
\end{equation}

Then, the first term in (\ref{eq:03}) describes the pure gravity
while the second one  is the particle term. More precisely, $h$
includes the translation and Lorentz transformation acting on the
particle and providing it with an arbitrary position, momentum, and
angular momentum, while ${\cal C}$ is a gauge algebra element
characterizing the particle at rest at the origin,
\begin{equation}\label{eq:04b}
{\cal C} = \frac{m}{\sqrt\Lambda}\, J_0 + s\, P_0\,,
\end{equation}
where $m$ is the mass and $s$ the spin of the particle in the rest
frame.

Last but not least, the integrand in the third term in (\ref{eq:03})
can be seen as a constraint relating the curvature with the mass/spin
of the particle
\begin{equation}\label{eq:05}
\frac{k}{2\pi} \F_S = h {\cal C} h^{-1} \delta^2(\vec{x})\,
dx^1 \wedge dx^2\,,
\end{equation}
enforced by the Lagrange multiplier $A_0$. It bounds together the
gravitational and particle degrees of freedom, therefore we may use
it to solve for the latter in terms of the former.

In order to proceed further we divide the space into two regions:
the disc ${\cal D}$ with  the particle at its center, on which we
introduce the coordinates $r \in [0,1]$, $\phi \in [0,2\pi]$, and the
asymptotic empty region ${\cal E}$ (with $r \geq 1$). They share the
boundary $\Gamma$ at $r = 1$. Then by virtue of (\ref{eq:05}) in the
asymptotic region the connection is flat and has the form
\begin{equation}\label{eq:06}
\A_S^{({\cal D})} = \gamma d\gamma^{-1}\,,
\end{equation}
where $\gamma$ is an element of the gauge group. In the particle
region ${\cal D}$ the general solution of (\ref{eq:05}) can also be
found and it is given by
\begin{equation}\label{eq:07}
\A_S^{({\cal E})} = \bar\gamma\, \frac{1}{k} {\cal C} d\phi\,
\bar\gamma^{-1} + \bar\gamma d\bar\gamma^{-1}\,, \quad
\bar\gamma(0) = h\,,
\end{equation}
(note that $dd\phi = 2\pi\, \delta^2(\vec{x})\, dx^1 \wedge dx^2$).
Substitution  of (\ref{eq:06}) into  the Lagrangian yields the
boundary term and the so-called WZW term. The same is the case for
the second term in (\ref{eq:07}). Contribution from the first term
in (\ref{eq:07}) can be rewritten as
\begin{equation}\label{eq:08}
2\left<\partial_0 (\bar\gamma {\cal C} \bar\gamma^{-1} d\phi) \wedge
\bar\gamma d\bar\gamma^{-1}\right> = -2\partial_0
\left<{\cal C} d\phi \wedge \bar\gamma^{-1} d\bar\gamma\right> -
2d\left<{\cal C} d\phi\, \bar\gamma^{-1} \dot{\bar\gamma}\right> +
4\pi\, \delta(\vec{x})\, dx^1\wedge dx^2
\left<{\cal C}\, \bar\gamma^{-1} \dot{\bar\gamma}\right>\,,
\end{equation}
where the first term can be neglected being a total time derivative
and the last one cancels the particle term in (\ref{eq:03}). Summing
all the contributions and adopting the opposite orientation of the
boundary $\Gamma$ for the terms coming from the disc ${\cal D}$ we
obtain the total Lagrangian
\begin{align}
L & = \frac{k}{4\pi} \int_{\Gamma} \left<\dot\gamma \gamma^{-1} d
\gamma \gamma^{-1} - \dot{\bar\gamma} \bar\gamma^{-1} d \bar\gamma
\bar\gamma^{-1} + \frac{2}{k} {\cal C} d\phi\, \dot{\bar\gamma}
\bar\gamma^{-1}\right> \nonumber\\
 & + \frac{k}{4\pi} \int_{\cal E}
\left<\dot\gamma \gamma^{-1} d \gamma\gamma^{-1} \wedge d \gamma
\gamma^{-1}\right> + \frac{k}{4\pi} \int_{\cal D}
\left<\dot{\bar\gamma} \bar\gamma^{-1} d \bar\gamma \bar\gamma^{-1}
\wedge d \bar\gamma \bar\gamma^{-1}\right>\,.
\label{eq:09}\end{align}
In the next step we impose the continuity condition on the boundary
$\Gamma$, $\A_S^{({\cal D})}|_\Gamma = \A_S^{({\cal E})}|_\Gamma$.
Solving this equation we find the expression
\begin{equation}\label{eq:10}
\gamma^{-1}|_\Gamma = N e^{\frac{1}{k} {\cal C} \phi}
\bar\gamma^{-1}|_\Gamma\,, \quad dN = 0\,,
\end{equation}
where $N = N(t)$ is an arbitrary gauge group element.

The idea now is to use the continuity condition (\ref{eq:10}) to
simplify the Lagrangian (\ref{eq:09}). Unfortunately, this condition
is very difficult to disentangle in the case of the gauge group
$\SO(3,1)$. Therefore in what follows we will consider only
contractions of this gauge group, leading to the effective gauge
group having the form of the semidirect product of some new group
and the dual of its algebra $G \ltimes \mathfrak{g}^*$, see
\cite{Meusburger:2003hc}.

Before deriving the main result of this paper, let us pause for a
moment to recall the well known construction in the case of the
standard contraction of de Sitter group $\SO(3,1)$ to the
Poincar\`{e} group, which can be presented as
$\SO(2,1) \ltimes \so(2,1)^* \simeq \SO(2,1) \ltimes \mathbbm{R}^3$.
To this end we introduce the rescaled translation generators,
$\tilde P_a \equiv \sqrt{\Lambda} P_a$. Then (\ref{eq:01}) is
replaced by
\begin{equation}\label{eq:11}
[J_a,J_b] = \epsilon_{ab}^{\ \ c}\, J_c\,, \quad [J_a,\tilde P_b] =
\epsilon_{ab}^{\ \ c}\, \tilde P_c\,, \quad [\tilde P_a,\tilde P_b]
= -\Lambda \epsilon_{ab}^{\ \ c}\, J_c\,
\end{equation}
and $\langle J_a \tilde P_b\rangle = \sqrt\Lambda\, \eta_{ab}$. If we
now take the limit $\Lambda \rightarrow 0$ then
$[\tilde P_a,\tilde P_b] = 0$, while the remaining commutators are
unchanged. Moreover, $\sqrt\Lambda$ in the scalar product cancels its
inverse in the definition of $k/4\pi$, cf.\ (\ref{eq:04a}) and thus
no divergencies appear in this limit. Furthermore, with the help of
Cartan decomposition a gauge group element can be written as a
product
\begin{equation}\label{eq:12}
g = \mathfrak{j}\, \mathfrak{p} = (\iota_3 + \iota^a J_a)
(1 + \xi^a \tilde P_a)\,, \quad \mathfrak{j} \in \SO(2,1)\,, \quad
\mathfrak{p} \in \so(2,1)^*\,,
\end{equation}
where the coordinates $\iota$ on $\SO(2,1)$ group satisfy
$\iota_3^2 + \frac{1}{4} \iota_a \iota^a = 1$.

Applying (\ref{eq:12}) to group elements in the Lagrangian
(\ref{eq:09}) we find that the WZW terms cancel out and only the
boundary ones remain, giving
\begin{equation}\label{eq:13}
L = \frac{k}{2\pi} \int_{\Gamma} \left<\mathfrak{j}^{-1}
\dot{\mathfrak{j}}\, d\xi - \bar{\mathfrak{j}}^{-1}
\dot{\bar{\mathfrak{j}}}\, d\bar\xi + \frac{1}{k} {\cal C}_P d\phi\, \bar{\mathfrak{j}}^{-1} \dot{\bar{\mathfrak{j}}} +
\frac{1}{k} {\cal C}_J d\phi\, \left[\bar{\mathfrak{j}}^{-1}
\dot{\bar{\mathfrak{j}}}, \bar\xi\right]\right>\,,
\end{equation}
where we also neglected the total time derivative $\frac{1}{k}\,
{\cal C} d\phi\, \dot{\bar\xi}$. Meanwhile, the sewing condition
(\ref{eq:10}) factorizes into $\mathfrak{j}^{-1} = \mathfrak{n}\,
e^{\frac{1}{k} {\cal C}_J \phi} \bar{\mathfrak{j}}^{-1}$ and $-\xi =
\Ad(\mathfrak{n})\, h - \Ad(\mathfrak{n}\,
e^{\frac{1}{k} {\cal C}_J \phi})\, \bar\xi$,
where we write the decomposition (\ref{eq:12}) of $N$ as $N =
\mathfrak{n}\, (1 + h)$, $\mathfrak{n} \in \SO(2,1)$,
$h \in \so(2,1)^*$. ${\cal C}$ can be separated into ${\cal C} =
{\cal C}_J + {\cal C}_P$, ${\cal C}_J = m/\sqrt\Lambda\, J_0$,
${\cal C}_P = s/\sqrt\Lambda\, \tilde P_0$. Substituting the above
expressions into (\ref{eq:13}) we get
\begin{equation}\label{eq:14}
L = \frac{k}{2\pi} \int_\Gamma
d \left<e^{-\frac{1}{k} {\cal C}_J \phi} \dot{\mathfrak{n}}^{-1}
\mathfrak{n}\, e^{\frac{1}{k} {\cal C}_J \phi} \bar{\xi} -
\mathfrak{n}^{-1} \dot{\mathfrak{n}}\,
\frac{1}{k} {\cal C}_P \phi\right>\,,
\end{equation}
where $\bar{\xi} \equiv \bar{\xi}^a \tilde P_a$. Integrating
(\ref{eq:14}) over $\phi$ from $0$ to $2\pi$ and noticing that
$\bar\xi$ is a single-valued function on $\Gamma$, hence
$\bar\xi(0) = \bar\xi(2\pi)$, we finally obtain
\cite{Meusburger:2003ta}
\begin{equation}\label{eq:15}
L = \kappa \,x^a\, \left(\dot\Pi\, \Pi^{-1}\right)_a -
s \left(\mathfrak{n}^{-1} \dot{\mathfrak{n}}\right)_0\,,
\end{equation}
with the new variables of particle's position
$x = x^a\, \tilde P_a \equiv \mathfrak{n}\, \bar\xi(0)\,
\mathfrak{n}^{-1}$ and ``group valued momentum'' $\Pi$, defined as
\begin{equation}\label{eq:16}
\Pi \equiv \mathfrak{n}\, e^{-\frac{2\pi}{k} {\cal C}_J}
\mathfrak{n}^{-1} = e^{-\frac{m}{\kappa} \mathfrak{n} J_0
\mathfrak{n}^{-1}}\,.
\end{equation}
Thus the Lagrangian (\ref{eq:15}) describes a deformed particle,
whose momentum is now given by the group element $\Pi$ defined
above instead of the algebra element $m J_0$.

The group valued momentum $\Pi$ is not arbitrary, but is given by
the conjugation of $e^{-\frac{2\pi}{k} {\cal C}_J}$ by the Lorentz
group element $\mathfrak{n}$. If we parametrize
$m\, \mathfrak{n} J_0 \mathfrak{n}^{-1} = q^a J_a$ then from
(\ref{eq:16}) we find
\begin{equation}\label{eq:16a}
\Pi = p_3 - \frac{1}{\kappa}\, p^a J_a\,, \quad p_3^2 +
\frac{1}{4\kappa^2}\, p_ap^a = 1\,, \quad p_3 =
\cos\left(\frac{|q|}{2\kappa}\right)\,, \quad p^a =
2\kappa \frac{q^a}{|q|}\, \sin\left(\frac{|q|}{2\kappa}\right)\,.
\end{equation}
It follows that the momenta $p_a$ are coordinates on the three
dimensional Anti-de Sitter space constrained by the deformed mass
shell condition. Introducing the Lagrange multiplier $\lambda$, the
final action for the spinless case ${\cal C}_P = 0$ can be written
in the components as
\begin{equation}\label{eq:17}
S = \int\! dt \left(p_3 \dot p_a x^a +
\frac{1}{2\kappa} \epsilon_{abc}\, \dot p^a x^b p^c -
\dot p_3 p_a x^a\right) + \lambda \left(p_ap^a -
4\kappa^2 \sin^2\frac{m}{2\kappa}\right)\,,
\end{equation}
where $p_3 \equiv \sqrt{1 - \frac{1}{4\kappa^2}\, p_ap^a}$. The
detailed discussion of the properties of this action can be found in
\cite{Matschull:1997du}. In the spinning case there is an additional
term of the form $-s\, (n_3 \dot n_0 - n_0 \dot n_3 +
\frac{1}{2} (n^1 \dot n^2 - n^2 \dot n^1))$.

Let us now turn to the main result of this paper. The contraction of
the de Sitter group $\SO(3,1)$ to Poincar\'e group is pretty well
known and the resulting Lagrangian (\ref{eq:15}) has been derived
and thoroughly analyzed  e.g., in \cite{Matschull:1997du},
\cite{Meusburger:2003ta}. It turns out, however, that there exist
another contraction of the de Sitter group that to our knowledge has
not been discussed in the literature. Contrary to the case
considered above, where the translation sector of $\SO(3,1)$ was
``flattened'', we consider ``flattening'' of the Lorentz sector of
$\SO(3,1)$.

To describe this new contraction let us return to the original
algebra (\ref{eq:01}) and consider its Iwasawa decomposition into
$\SO(2,1)$ and $\AN(2)$. The generators of the latter are defined as
a linear combination of the original Lorentz and translation
generators
\begin{equation}\label{eq:aa}
S_a =  P_a + \epsilon_{a0b}\, J^b\,,
\end{equation}
so that we have
\begin{equation}\label{eq:aa1}
[J_a,J_b] = \epsilon_{ab}^{\ \ c}\, J_c\,, \quad [J_a,S_b] =
\epsilon_{ab}^{\ \ c}\, S_c - \eta_{ab}\, J_0 + \eta_{b0}\, J_a\,,
\quad [S_a,S_b] = \eta_{a0}\, S_b - \eta_{b0}\, S_a\,.
\end{equation}
The virtue of this decomposition is that the generators $J_a$ and
$S_a$ form subalgebras of the algebra $\so(3,1)$; the price to pay,
however, is that the  cross commutators become quite complicated.
Let us now rescale $\tilde J_a \equiv \sqrt{\Lambda} J_a$ to obtain
\begin{equation}\label{eq:aa2}
[\tilde J_a,\tilde J_b] = \sqrt\Lambda\, \epsilon_{abc}\,
\tilde J^c\,, \quad [\tilde J_a, S_b] = \sqrt\Lambda\,
\epsilon_{abc}\, S^c + (\eta_{b0}\, \tilde  J_a -
\eta_{ab}\, \tilde J_0)\,, \quad [S_a,S_b] = \eta_{a0}\, S_b -
\eta_{b0}\, S_a\,,
\end{equation}
which after contraction $\Lambda \rightarrow 0$ takes the form
\begin{equation}\label{eq:bb}
[\tilde J_a,\tilde J_b] = 0\,, \quad [\tilde J_a, S_b] =
 (\eta_{b0}\, \tilde  J_a - \eta_{ab}\, \tilde  J_0)\,,
\quad [S_a,S_b] = \eta_{a0}\, S_b - \eta_{b0}\, S_a\,.
\end{equation}
It is worth mentioning that, as it was in the case of the Poicar\'e
algebra above, the algebra (\ref{eq:bb}) is a Lie algebra of the
group  $G \ltimes \mathfrak{g}^*$, where $G$ is now the group
$\AN(2)$ generated by the last commutator in (\ref{eq:bb}).

In terms of the new generators the scalar products read
\begin{equation}\label{eq:cc}
\left<\tilde J_a S_b\right> =\sqrt\Lambda\, \eta_{ab}\,, \quad
\left<\tilde J_a \tilde J_b\right> = \left< S_a S_b\right> = 0\,.
\end{equation}
In spite of the fact that this scalar product becomes degenerate in
the limit $\Lambda \rightarrow 0$, in the effective particle action
$\Lambda$ is cancelled out and the contraction limit is not singular.

A gauge group element can now be decomposed into
\begin{equation}\label{eq:dd}
\gamma = \mathfrak{j}\, \mathfrak{s} = (1 + \iota^a \tilde J_a)\,
e^{\sigma^i\, S_i}\, e^{\sigma^0\, S_0}\,, \quad i = 1,2\,,
\end{equation}
where for $\mathfrak{s}$ we use the parametrization that proved
convenient in the context $\kappa$-Poincar\'{e} theories
\cite{Kowalski-Glikman:2013rxa} and is related to the other
parametrization $\mathfrak{s} = \xi_3 + \xi^a S_a$ via $\sigma^0 =
2\log(\xi_3 + \frac{1}{2} \xi^0)$, $\sigma^i = (\xi_3 +
\frac{1}{2} \xi^0)\, \xi^i$.

Since it is our goal to obtain a curved momentum space after the
$\Lambda \rightarrow 0$ limit is taken, we must change the form of
${\cal C} = {\cal C}_J + {\cal C}_S$, describing the
particle at rest, so as to have the mass in the $S$ sector.
Adjusting dimensions properly we get ${\cal C}_J = s/\sqrt\Lambda\,
\tilde J_0$, ${\cal C}_S = m/\sqrt\Lambda\, S_0$.

After these preparations we can return to the formulae
(\ref{eq:09}), (\ref{eq:10}). Writing $N = (1 + n)\, \mathfrak{h}$,
with $\mathfrak{h}\in\AN(2)$, $n\in\so(2,1)$ and using the
factorization (\ref{eq:dd}) one first finds that
\begin{equation}\label{eq:22}
\mathfrak{s}^{-1} = \mathfrak{h}
\exp\left(\frac{1}{k} {\cal C}_S \phi\right)
\bar{\mathfrak{s}}^{-1}\,.
\end{equation}
Next, from the commutation relations (\ref{eq:bb}), for an arbitrary
$\mathfrak{s}$ we have
\begin{equation}\label{eq:23}
\mathfrak{s} \exp\left(\frac{1}{k} {\cal C}_J \phi\right)
\mathfrak{s}^{-1} = \exp\left(\frac{1}{k} {\cal C}_J \phi\right)\,.
\end{equation}
As a result one obtains the second condition
\begin{equation}\label{eq:24}
\mathfrak{u} = \exp\left(\frac{1}{k} {\cal C}_J \phi\right)
\mathfrak{s}\, (1 - n) \mathfrak{s}^{-1}\,,
\end{equation}
where we denote $\mathfrak{u} \equiv
\bar{\mathfrak{j}}^{-1} \mathfrak{j}$.

We now plug the factorization (\ref{eq:dd}) into our starting Lagrangian
(\ref{eq:09}), finding
\begin{equation}\label{eq:25}
L = \frac{k}{2\pi} \int_\Gamma \left<\dot{\mathfrak{u}}\,
\mathfrak{u}^{-1} \left(d\bar{\mathfrak{s}}\,
\bar{\mathfrak{s}}^{-1} - \bar{\mathfrak{s}}\,
\frac{1}{k}\, {\cal C} d\phi\, \bar{\mathfrak{s}}^{-1}\right) +
\frac{1}{k}\, {\cal C} d\phi\, \bar{\mathfrak{s}}^{-1}
\dot{\bar{\mathfrak{s}}}\right>\,.
\end{equation}
Substituting the continuity conditions for $\mathfrak{u}$ and
$\mathfrak{s}$ into (\ref{eq:25}) and keeping in mind the limit
$\Lambda \rightarrow 0$ we obtain
\begin{equation}\label{eq:26}
L = \frac{k}{2\pi} \int \left<\partial_0 \left(\bar{\mathfrak{s}}
e^{-\frac{1}{k} {\cal C}_S \phi} \mathfrak{h}^{-1} n \mathfrak{h}
e^{\frac{1}{k} {\cal C}_S \phi} \bar{\mathfrak{s}}^{-1}\right)
\left(-d\bar{\mathfrak{s}}\, \bar{\mathfrak{s}}^{-1} +
\bar{\mathfrak{s}}\, \frac{1}{k}\, {\cal C}_S d\phi\,
\bar{\mathfrak{s}}^{-1}\right) + \frac{1}{k}\, {\cal C}_J d\phi\,
\bar{\mathfrak{s}}^{-1} \dot{\bar{\mathfrak{s}}}\right>\,.
\end{equation}
In the last step, neglecting the total time derivative and
integrating over the angular variable we eventually obtain the
expression very similar to (\ref{eq:15})
\begin{equation}\label{eq:31}
L = \frac{k}{2\pi} \left<\Pi\, \dot\Pi^{-1}\, x\right> +
\left<{\cal C}_J \bar{\mathfrak{s}}^{-1}
\dot{\bar{\mathfrak{s}}}\right>\,,
\end{equation}
but with the deformed momenta $\Pi$ being now the elements of the
$\AN(2)$ group, instead of the Lorentz group $\SO(2,1)$, to wit
\begin{equation}\label{eq:32}
\Pi \equiv \bar{\mathfrak{s}}\, e^{\frac{2\pi}{k} {\cal C}_S}
\bar{\mathfrak{s}}^{-1}\,, \qquad x \equiv \bar{\mathfrak{s}}\,
\mathfrak{h}^{-1}(0) n(0) \mathfrak{h}(0) \bar{\mathfrak{s}}^{-1}\,.
\end{equation}
The Lagrangian (\ref{eq:31}) is the main result of our paper.

Since the Lorentz group $\SO(2,1)$ is, as a manifold, the three
dimensional Anti-de Sitter space, while $\AN(2)$ is a submanifold of
the three dimensional de Sitter space, we managed to obtain the
momentum space of positive, instead of the negative, constant
curvature. Moreover, contrary to (\ref{eq:15}), which is defined
only in 2+1 dimension, the expression (\ref{eq:31}) can be readily
generalized to any spacetime dimension.

Let us now turn to the detailed discussion of the properties of
Lagrangian (\ref{eq:31}). The first thing to notice is that the
first equation in (\ref{eq:32}) puts severe restrictions on the form
of momentum $\Pi$. Indeed if we write
\begin{equation}\label{eq:33}
\Pi = e^{p^i/\kappa\, S_i}\, e^{p^0/\kappa\, S_0}\,,
\end{equation}
and take $\bar{\mathfrak{s}}$ to have the form
\begin{equation}\label{eq:33a}
\bar{\mathfrak{s}} = e^{\bar\sigma^i\, S_i}\,
e^{\bar\sigma^0\, S_0}\,,
\end{equation}
we immediately find that
\begin{equation}\label{eq:34}
p^0 = m\,, \qquad p^i = \kappa\, (1 - e^{\frac{m}{\kappa}})\,
\bar\sigma^i\,.
\end{equation}
As it was in the case considered above these equations play a role
of the mass shell relation, and force the energy to be constant,
independently of the particle dynamics. In the undeformed case
(which can be obtained in the limit $\kappa\rightarrow\infty$) such
mass shell condition makes the particle effectively frozen, it can
not move, and for that reason, following \cite{Bergshoeff:2014jla}
we call it the ``Carroll particle.''

In the following discussion we will consider only the spinless case.
It turns out that this is in fact the general case, because the
spin term does not contribute nontrivially to the equations of
motion. Indeed an arbitrary variation
$\delta \bar{\mathfrak{s}} = \varpi\, \bar{\mathfrak{s}}$,
$\delta \bar{\mathfrak{s}}^{-1} = -\bar{\mathfrak{s}}^{-1} \varpi$ of
this term results in a total time derivative
\begin{equation}\label{eq:34a}
\delta\left<{\cal C}_J \bar{\mathfrak{s}}^{-1}
\dot{\bar{\mathfrak{s}}}\right> = \left<{\bar{\mathfrak{s}}}
{\cal C}_J \bar{\mathfrak{s}}^{-1} \dot\varpi\right> =
\frac{d}{dt}\left<{\cal C}_J \varpi\right>\,,
\end{equation}
because $\tilde J_0$ commutes with all $S$ generators
(cf.\ (\ref{eq:bb}).)

Let us discuss the deformed Lagrangian in more details. Expressed in
components of momentum $p_a$ it reads
\begin{equation}\label{eq:35}
L = x^0\, \dot p_0 + x^i\, \dot p_i - \kappa^{-1} x^i\, p_i\, \dot
p_0 + \lambda (p_0^2 - m^2)\,,
\end{equation}
where, as before, we introduce the Lagrange multiplier $\lambda$ to
enforce the mass shell constraint. The equations of motion following
from variations over $x$ are momentum conservations $\dot p_a = 0$,
while the ones resulting from the variation over momenta give
\begin{equation}\label{eq:35a}
\dot x^0 = 2\lambda\, p_0 = 2\lambda\, m\,, \quad \dot x^i = 0\,,
\end{equation}
so that indeed the Carroll particle is always at rest. Furthermore,
from (\ref{eq:32}) we may also find the explicit expressions for the
components $x^0 = n^0 - (e^{\zeta_0 - \bar\sigma_0} \bar\sigma^i -
\zeta^i)\, n_i$, $x^i = e^{\zeta_0 - \bar\sigma_0} n^i$. Then
(\ref{eq:35a}) will give some conditions for coordinates
of $n$, $\mathfrak{h} = e^{\zeta^i S_i} e^{\zeta^0 S_0}$ and
$\bar{\mathfrak{s}}$.

The presence of the nonlinear, deformed term in the Lagrangian
(\ref{eq:35}) results in the nontrivial Poisson bracket algebra of
the $\kappa$-deformed phase space \cite{AmelinoCamelia:1997jx}
\begin{equation}\label{eq:35b}
\left\{x^i,p_j\right\} = \delta^i_j\,, \quad
\left\{x^0,p_0\right\} = 1\,, \quad
\left\{x^0,p_i\right\} = -\frac{1}{\kappa}\, p_i\,, \quad
\left\{x^0,x^i\right\} = \frac{1}{\kappa}\, x^i\,,
\end{equation}

Meanwhile, symmetries of the action obtained from the Lagrangian
(\ref{eq:35}) form the algebra of infinitesimal deformed Carroll
transformations, containing
\begin{itemize}
\item rotations
\begin{equation}\label{eq:36}
\delta x^i = \rho\, \epsilon^i{}_j\, x^j\,, \quad \delta p_i = \rho\, \epsilon_i{}^j\,
p_j\,, \quad \delta x^0 = \delta p_0 = 0\,;
\end{equation}
\item deformed boosts
\begin{equation}\label{eq:37}
\delta x^0 = (1 + \kappa^{-1} p_0)\, \lambda_i\, x^i\,, \quad
\delta p_i = -\lambda_i\, p_0\,,
\quad \delta x^i = \delta p_0 = 0\,;
\end{equation}
\item deformed translations
\begin{equation}\label{eq:38}
\delta x^0 = a^0\,, \quad \delta x^i = e^{p_0/\kappa}\, a^i\,, \quad
\delta p_a = 0\,;
\end{equation}
\item the spatial conformal transformation
\begin{equation}\label{eq:39}
\delta x^i = \eta\,  x^i\,, \quad \delta p_i = -\eta\,
p_i\,, \quad \delta x^0 = \delta p_0 = 0\,,
\end{equation}
\end{itemize}
where $\rho$, $\lambda_i$, $a^a$, $\eta$ are parameters of the respective transformations.

Actually, as noted in \cite{Bergshoeff:2014jla} the undeformed
Carroll particle has an infinite dimensional symmetry. This property
holds in the deformed case as well, and the generator of the
infinitesimal transformations $\delta\phi^a =
\left\{\phi^a,G\right\}$, where $\phi$ is an arbitrary function on
phase space, is given by
\begin{equation}\label{eq:40}
G = f(p_0/\kappa) p_0\, \xi^0(x^i) + p_i\, \xi^i(x^i)\,, \quad
f(0) = 1\,,
\end{equation}
where $f(p_0/\kappa)$ is an arbitrary function of energy, while
$\xi^i(x^i)$, $\xi^0(x^i)$ are arbitrary functions of position.

Let us complete this paper with some comments.

First it should be stressed that the particle model we derived
(\ref{eq:31}) can be extended to any number of spacetime dimensions
$D+1$, simply by replacing the group $\AN(2)$ with $\AN(D)$. This
makes it potentially much more relevant for real physical systems
than the model based on Poincar\'{e} group (\ref{eq:15}), whose
application is strictly restricted to 2+1 dimensions.

In our view, the most significant result of this paper is the
derivation of a particle model with $\kappa$-deformed phase space
(\ref{eq:35b}) from the first principles as a deformation of the
free particle model resulting from the interaction of the particle
with its own gravitational field. However, it turns out that the
model we obtained is not the $\kappa$-Poincar\'{e} particle discussed
in the context of Doubly Special Relativity
\cite{KowalskiGlikman:2003we} or Relative Locality
\cite{AmelinoCamelia:2011bm}, \cite{Kowalski-Glikman:2013rxa}, but
the deformed Carroll particle, with completely frozen dynamics. It
is therefore still an open problem if the $\kappa$-Poincar\'{e}
particle can be derived from the particle-gravity system as an
effective deformed particle theory. We will revisit this issue in the
forthcoming paper.

There is a curious similarity between Carrollian relativity, being
the relativistic theory obtained in the limit $c \rightarrow 0$, in
which no local interactions are possible and the very similar
feature of the 2+1 gravity and the topological limit of gravity in
3+1 dimensions \cite{Freidel:2005ak}, \cite{Freidel:2006hv},
\cite{KowalskiGlikman:2008fj}. Especially in the 3+1 case it would be
of interest to find out if gravity is described by a topological
field theory in the Carrollian limit (for some discussion of this
issue see \cite{Dautcourt:1997hb}.)

As already mentioned, the Carrollian limit appears in many distinct
areas of theoretical physics, the common feature of whose is the
presence of gravity in one form or another. The free and interacting
particle models are extremely useful in that they help to grasp the
underlying physics. It seems that the deformed model presented here,
which already takes into account self-gravitational interactions
might be of great interest, especially in the context of
cosmological investigations \cite{Mielczarek:2012tn}.

\section*{Acknowledgment} This work was supported by funds provided
by the National Science Center under the agreement
DEC-2011/02/A/ST2/00294. TT acknowledges the support by the
Foundation for Polish Science International PhD Projects Programme
co-financed by the EU European Regional Development Fund and the
additional funds by the European Human Capital Program.

\end{document}